\documentclass[aps,longbibliography,showpacs,twocolumn,superscriptaddress,amsmath,amssymb,verbatim]{revtex4-1}
\usepackage{multirow,array}
\usepackage{siunitx}
\usepackage{balance}
\usepackage[thinlines]{easytable}
\usepackage{graphicx}% Include figure files
\usepackage{lmodern}
\usepackage{epstopdf}
\usepackage{dcolumn}% Align table columns on decimal point
\usepackage{bm}% bold math
\usepackage{subfigure}
\usepackage{makecell}
\usepackage{amsmath} %math
\usepackage[pdftex,colorlinks=true,citecolor=blue,linkcolor=blue,urlcolor=blue,bookmarks=true]{hyperref}
\usepackage{booktabs}
\usepackage{multirow} % Add this package for multirow functionality
\usepackage{array} % For better array manipulation

\usepackage{color}
\usepackage{textcomp}
\definecolor{red}{rgb}{1,0,0}

\definecolor{blue}{rgb}{0,0,1}

\definecolor{green}{rgb}{0,1,0}

\usepackage{multirow}
\usepackage[version=4]{mhchem}
\usepackage[utf8]{inputenc}
\begin{document}
	\preprint{APS}

\title{Multistage  spin correlations in the $s$ = 1/2 stuffed hyper-star lattice Li$_{2}$Cu$_{2}$(MoO$_{4}$)$_{3}$ 
 }

\author{J. Khatua}
\affiliation{Department of Physics, Sungkyunkwan University, Suwon 16419, Republic of Korea}
\author{Taeyun Kim}
\affiliation{Department of Physics, Sungkyunkwan University, Suwon 16419, Republic of Korea}
\author{G. Senthil Murugan}
\homepage{nanosen@gmail.com}
\affiliation{Institute of Physics, Academia Sinica, Taipei 11529, Taiwan}	
\affiliation{Department of Physics, St. Joseph's College of Engineering, OMR, Chennai 600 119, India}
\author{S. M. Kumawat}
\affiliation{Department of Physics and Center for Quantum Frontiers of Research and Technology (QFort), National Cheng Kung University, Tainan 70101, Taiwan}
  \author{C.-L. Huang} 
\affiliation{Department of Physics and Center for Quantum Frontiers of Research and Technology (QFort), National Cheng Kung University, Tainan 70101, Taiwan}
\author{Yugo Oshima}
\affiliation{RIKEN Pioneering Research Institute, Wako, Saitama 351-0198, Japan}
\author{Hiroyuki Nojiri}
\affiliation{Institute for Materials Research, Tohoku University, Sendai 980-8577, Japan}
\author{Gerald Morris}
\affiliation{Centre for Molecular and Materials Science, TRIUMF, Vancouver, British Columbia V6T 2A3, Canada}
\author{ Sarah R. Dunsiger}
\affiliation{Centre for Molecular and Materials Science, TRIUMF, Vancouver, British Columbia V6T 2A3, Canada}
\author{Heung-Sik Kim}
\affiliation{Department of Energy Technology, Korea Institute of Energy Technology, Naju-si 58217, Republic of Korea}
\author{K. Sritharan}
\affiliation{Institute of Physics, Academia Sinica, Taipei 11529, Taiwan}
\author{Shankar Mani}
\affiliation{Institute of Physics, Academia Sinica, Taipei 11529, Taiwan}
     \author{R. Sankar}
\homepage{sankarndf@gmail.com}
\affiliation{Institute of Physics, Academia Sinica, Taipei 11529, Taiwan}
\author{Kwang-Yong Choi}
 \homepage{choisky99@skku.edu}
\affiliation{Department of Physics, Sungkyunkwan University, Suwon 16419, Republic of Korea}

\date{\today}

\begin{abstract}

Star lattice, which can be visualized as a honeycomb network with each vertex replaced by a triangle, provides a rare platform for realizing exotic quantum states such as quantum spin liquids and disorder-driven random-singlet (RS) states. Herein, we investigate the ground-state properties of the three-dimensional (3D) stuffed hyper-star lattice Li$_2$Cu$_2$(MoO$_4$)$_3$, which exhibits a crossover from short-range spin correlations to a disorder-driven RS-like state below $T^{*}\sim$15.8 K. Thermodynamic and microscopic measurements capture this crossover through a change in the power-law behavior of various observables, from $\sim T^{0.25}$ for $T > T^{*}$ to $\sim T^{-0.50}$ for $T < T^{*}$. Upon further cooling, a quasi-frozen state emerges near $T_{\rm f} = 0.32$ K, likely associated with weakly coupled spin chains within the hyper-star spin network. Our results underscore the crucial role of orphan spins and weak residual interactions in stabilizing a disorder-driven quantum-disordered state in 3D.

\end{abstract}
\maketitle
\section{Introduction}
Quantum spin systems provide an ideal platform for exploring exotic many-body quantum phenomena driven by strong quantum fluctuations that go beyond the conventional Landau paradigm of symmetry-breaking states~\cite{Balents2010,Savary2016}. One prominent example of a quantum spin-disorder ground state is the random-singlet (RS) state. Initially established in one-dimensional (1D) spin chains, the RS state emerges when quenched exchange randomness drives the system toward a disorder-dominated fixed point~\cite{PhysRevLett.43.1434,PhysRevB.22.1305}. In this regime, spins pair into singlets over a broad distribution of energy scales, forming a hierarchy of correlated singlets without breaking any global symmetry~\cite{PhysRevLett.43.1434}. Experimentally, this manifest as a universal power-law scaling in thermodynamic quantities, predicted by  renormalization group theory for infinitely disordered systems~\cite{IGLOI2005277,PhysRevLett.45.1303}. \\
In recent years, the search for RS states in higher-dimensional frustrated systems has gained renewed interest, particularly in the presence of quenched disorder arising from bond randomness or site dilution~\cite{PhysRevX.8.031028,Kimchi2018}. For instance, several 2D disorder-induced RS states have been reported, including the square-lattice compound Sr$_{2}$CuTe$_{1-x}$W$_{x}$O$_6$~\cite{PhysRevLett.126.037201,Mustonen2018} and the triangular-lattice system Y$_2$CuTiO$_6$~\cite{PhysRevLett.125.117206}. Similar to their 1D counterparts, these 2D RS analogs are characterized by a power-law low-energy density of states, $D(E) \propto E^{-\alpha}$, which gives rise to power-law dependencies in physical observables~\cite{PhysRevX.8.041040,PhysRevLett.127.127201,PhysRevLett.122.167202}.
However, the realization of RS phenomenology in 3D systems remains scarce. Theoretical studies suggest that in three dimensions, lattice vacancies not only lead to bond randomness but also generate emergent quasispins, localized magnetic moments associated with vacancy sites that coexist with the intrinsic bulk spins~\cite{PhysRevB.106.L140202}. Upon increasing vacancy concentration, the density of these quasispins grows and ultimately drive the system into a quantum disorder state~\cite{PhysRevB.106.L140202}.\\
The RS state exists in subtle competition with the quantum spin liquid (QSL) state. While a QSL is characterized by long-range quantum entanglement and fractionalized, deconfined excitations (e.g., spinons)~\cite{ANDERSON1973153,RevModPhys.89.025003,PhysRevB.65.165113,KHATUA20231}, the higher-dimensional RS state comprises a random network of short-range entangled singlets and emergent quasispins. Because both states lack conventional long-range order and can give rise to similar experimental signatures, distinguishing between them remains a central challenge in modern condensed matter physics~\cite{shimokawa202}. Systematic investigations of frustrated magnets in the presence of quenched disorder or site dilution are therefore crucial for elucidating the distinct physical behaviors of these two quantum-disordered states.\\ In this context, the star lattice, derived from a honeycomb lattice with each vertex replaced by a triangle, has emerged as a promising platform for frustration-driven quantum disorder state~\cite{PhysRevB.81.134418}. Unlike the bipartite honeycomb lattice, the star lattice is non-bipartite with coordination number $z$ = 3,  holding potential for hosting richer ground states than kagome systems~\cite{PhysRevB.70.174454}.
While theoretical studies predict a variety of quantum phases on the pristine star lattice~\cite{PhysRevB.92.155105,PhysRevLett.99.247203,PhysRevB.81.104429,PhysRevB.97.075146}, the introduction of bond randomness or vacancies is expected to promote RS phenomenology manifested by a power-law magnetic susceptibility, $\chi(T) \sim T^{\alpha'(T)-1}$, with $\alpha'(T) \rightarrow 0$ as $T \rightarrow 0$~\cite{PhysRevLett.127.127201}.\\ Motivated by this interplay between disorder and critical spin correlations, we investigate the ground-state properties and spin dynamics of Li$_2$Cu$_2$(MoO$_4$)$_3$, a stuffed 3D hyper-star lattice compound. Thermodynamic and local-probe resonance measurements establish Li$_2$Cu$_2$(MoO$_4$)$_3$ as a rare 3D frustrated magnet where Li/Cu site mixing induces a multi-stage development of spin correlations. At the higher-temperature regime, dominated by 3D random antiferromagnetic correlations, exhibits a broad heat-capacity maximum and power-law ($\sim T^{0.25}$)
behavior in both the muon spin relaxation rate and electron spin resonance (ESR) linewidth down to $T^{*}$ $\approx$ 15.8 K. Below $T^{*}$, emergent quasispins produce low-temperature power laws ($\sim T^{-0.50}$) in magnetic susceptibility, magnetic specific heat divided by temperature, and zero-field muon spin relaxation, indicative of the formation of an RS–like state. Eventually, a quasi-frozen state emerges below $T_{\rm f}$ = 0.32 K, likely linked to weakly coupled spin chains at the centers of the 3D hyper-star-lattice polygons.\\

\section{EXPERIMENTAL DETAILS}
Polycrystalline samples of Li$_2$Cu$_2$(MoO$_4$)$_3$ (LCMO)  were synthesized by a standard solid-state reaction route. High-purity precursor powders of Li$_2$CO$_3$, CuO, MoO$_3$ (with purity of 99.95$\%$) were weighed in stoichiometric proportions, thoroughly mixed, and finely ground using an agate mortar to ensure homogeneity. Prior to weighing, all reagents were preheated in air at 100$^\circ$C for 24 h to remove residual moisture. The homogenized mixture was initially calcined in air at 450$^\circ$C for 48 h in a ceramic crucible. To obtain a single-phase compound, the calcined powder was subjected to successive sintering steps at 500$^\circ$C and 550$^\circ$C for 48 h each, with intermediate grindings to enhance phase purity.
To confirm the phase purity and crystal structure, synchrotron X-ray diffraction (XRD) measurements were performed at the 09A beamline of NSRRC, Taiwan, at 120~K. The diffraction patterns were analyzed using the Rietveld refinement with the GSAS software package~\cite{Toby:hw0089}.\\
 Magnetic measurements, including dc magnetic susceptibility and isothermal magnetization, were carried out  using a superconducting quantum interference device vibrating-sample magnetometer (SQUID-VSM, Quantum Design, USA) in the temperature range of 2–300~K under magnetic fields up to 9~T. In addition, \textit{ac} magnetic susceptibilities were measured by the SQUID-VSM in the temperature range of 2–10~K at low frequencies ($\leq$ 950~Hz).  For measurements in the lower temperature range of 0.05–4~K, \textit{ac} magnetic susceptibility was measured using a Physical Property Measurement System (PPMS, Quantum Design, USA) equipped with a dilution refrigerator insert.\\
Pulsed-field magnetization measurements at 0.5~K were performed at the Institute for Materials Research, Japan, using pulsed magnetic fields with the induction method employing a pick-up coil device and a $^3$He cryostat.
Specific-heat measurements were recorded using a standard relaxation method in a PPMS in the the temperature range 0.2~K $\leq T \leq$ 300~K in several magnetic fields up to 9~T.\\
ESR experiments were conducted at RIKEN using a conventional X-band spectrometer (JEOL, JES-RE3X; $f = 9.12$~GHz).  A continuous-flow $^{4}$He cryostat was employed to maintain precise temperature control in the range of 3.8–60 K.\\
Muon spin relaxation ($\mu$SR) measurements were conducted at the M20 beamline of TRIUMF (Vancouver, Canada) under zero-field (ZF), longitudinal-field (LF), and weak-transverse-field (wTF) conditions. About 0.5\,g of powder samples was enclosed in a 2.5\,$\mu$m Mylar envelope (sample area $\sim$1.3\,$\times$\,1.3\,cm$^2$) and further wrapped in 50\,$\mu$m Scotch aluminum tape to minimize background signals from the copper holder. The assembly was fixed onto a copper fork sample stick and cooled using a $^4$He flow cryostat, providing stable temperature control between 2 and 120\,K. The time evolution of the muon spin asymmetry was analyzed using the Musrfit software package~\cite{SUTER201269}.\\
Density functional theory (DFT) calculations were performed using the OpenMX code within the generalized gradient approximation of Perdew–Burke–Ernzerhof (GGA-PBE) exchange–correlation functional, incorporating an on-site Hubbard \textit{U} = 6 eV to account for Cu 3\textit{d} correlations. The self-consistent results were subsequently used in the $J_{X}$ code to evaluate the exchange coupling parameters $J_{ij}^{\rm GGA}$ through the Green’s-function-based Liechtenstein approach~\cite{YOON2020106927}.
\begin{table} [b]
	\caption{\label{SIF}\label{TabRF} Atomic coordinates of Li$_{2}$Cu$_{2}$(MoO$_{4}$)$_{3}$ from a  Rietveld refinement of synchrotron x-ray diffraction data at 120 K. ($\chi^{2}$ = 4.57, Space group: $P 2_{1}/c,$ $a$ = 10.47 \AA , $b$ = 5.01 \AA, $c$ = 20.37 \AA, $ $ $\alpha$ = $\gamma = 90^{\circ}$, and $ \beta$ = 120.94$^{\circ}$.)} 
	\begin{tabular}{c c c c c  c c} % <-- Alignments: l for left, c for center, and r for right, with vertical lines in between
		\hline \hline
		Atom & Wyckoff position & \textit{x} & \textit{y} &\textit{ z}& Occ.\\
		\hline 
		Mo$_{1}$ & 4$e$ & 0.315 \ \ & 0.713 \ \ & 0.344\ \ & 1 \\
		Mo$_{2}$ & 4$e$ & 0.871 \ \ & 0.718 \ \ & 0.344\ \ & 1 \\
		Mo$_{3}$ & 4$e$ & 0.692 \ \ & 1.224 \ \ & 0.442\ \ & 1 \\
		Cu$_{1}$ & 4$e$ & 0.074 \ \ & 0.206 \ \ & 0.323\ \ & 0.30 \\
		Li$_{1}$ & 4$e$ & 0.074 \ \ & 0.206 \ \ & 0.323\ \ & 0.70 \\
		Cu$_{2}$ & 4$e$ & 0.953 \ \ & 0.751 \ \ & 0.525\ \ & 0.46 \\
		Li$_{2}$ & 4$e$ & 0.953 \ \ & 0.751 \ \ & 0.525\ \ & 0.54 \\
		Cu$_{3}$ & 4$e$ & 0.598 \ \ & 0.752\ \ & 0.525\ \ & 0.49 \\
		Li$_{3}$ & 4$e$ & 0.598 \ \ & 0.752\ \ & 0.525\ \ & 0.51 \\
		Cu$_{4}$ & 4$e$ & 0.499 \ \ & 0.888\ \ & 0.249\ \ & 0.80 \\
		Li$_{4}$ & 4$e$ & 0.499 \ \ & 0.888\ \ & 0.249\ \ & 0.20 \\
		O$_{1}$ & 4$e$ & 0.915 & 0.916 & 0.425 & 1\\
		O$_{2}$ & 4$e$&  0.436 & 0.913 & 0.425 & 1 \\
		O$_{3}$ & 4$e$& 0.258 & 0.430 & 0.374 & 1\\
		O$_{4}$ &  4$e$& 0.682 & 0.642 & 0.294 & 1\\
		O$_{5}$ & 4$e$& 0.407 & 0.639 & 0.294& 1\\	
		O$_{6}$ & 4$e$& 0.9912 & 0.432& 0.374& 1\\	
		O$_{7}$ & 4$e$& 0.164 & 0.931& 0.288& 1\\	
		O$_{8}$ & 4$e$& 0.759 & 0.955& 0.509& 1\\	
		O$_{9}$ & 4$e$&0.596 & 1.13& 0.346 & 1\\
		O$_{10}$ & 4$e$&0.912 & 0.931& 0.289 & 1\\	
		O$_{11}$ & 4$e$&0.846 & 1.428 & 0.460 & 1\\		
		O$_{12}$ & 4$e$&0.574 & 1.426 & 0.460 & 1\\			
		\hline
	\end{tabular}
	%\end{ruledtabular}
\end{table} 
\begin{table*}[t]
	\centering
	\caption{\label{tableangle}Cu-Cu distances, Cu–O–Cu bond angles, and  corresponding exchange interactions denoted as $J$ for Li$_{2}$Cu$_{2}$(MoO$_{4}$)$_{3}$. Although the absolute exchange values are not calculated, their expected magnitudes are compared with those reported for Cu-based compounds with bonding geometries.}
	\setlength{\tabcolsep}{6pt}\renewcommand{\arraystretch}{1.15}
	\begin{tabular}{|c|c|c|p{8.0cm}|}
		\hline
		Bond length (\AA) & Bond angle ($^\circ$) & $J$ & 
		\textbf{Expected range of $J$ (comparative notes).} \\ \hline
		
		Cu$1$–Cu$2$ = 3.22  & Cu$1$–O7–Cu$2$ = 106.3 & $J_{1}$ & 
		YCu$_3$(OH)$_{6.5}$Br$_{2.5}$: $\angle\text{Cu–O–Cu}$ =112-116$^{\circ}$, $J\sim60$ K, $d_{\rm Cu\!-\!Cu}\sim3.3$ \AA \ \ \cite{Chen_2025}. \\ \cline{1-3}
		
		Cu$1$–Cu$3$ = 3.22  & Cu$1$–O3–Cu$3$ = 106.5 & $J_{3}$ & \\ \cline{1-3}
		
		Cu$2$–Cu$3$ = 3.71 & Cu$2$–O8–Cu$3$ = 120.6 & $J_{2}$ &  YCu$_{3}$(OH)$_{6}$Cl$_{3}$: $\angle\mathrm{Cu\!-\!O\!-\!Cu} = 117.8^\circ$, $J$ = 87 K, 
		$d_{\rm Cu\!-\!Cu}\sim3.37$ \AA \ \ \cite{PhysRevLett.125.027203}.  \\ \hline
		
		Cu${4}$–Cu${4}$ = 2.50  & Cu${4}$–O9–Cu${4}$ = 73.5–75 & $J_{4}$ & 
		\multirow{6}{*}{\parbox[t]{8.0cm}{}} \\ \cline{1-3}
		
		Cu${1}$–Cu${1}$ = 3.60  & Cu${1}$–O7–Cu${1}$ = 101.4 & $J_{11}$ & Cu$_{2}$Cl(OH)$_{3}$: $\angle\mathrm{Cu\!-\!O\!-\!Cu} = 97.84^\circ$, $J\sim 1.3$ K, $d_{\rm Cu\!-\!Cu}\sim3.01$ \AA \ \ \cite{PhysRevLett.126.207201}. \\ \cline{1-3}
		Cu${2}$–Cu${2}$ = 3.05  & Cu${2}$–O11–Cu${2}$ = 94.5  & $J_{22}^{\rm a}$ & \\ \cline{1-3}
		Cu${2}$–Cu${2}$ = 3.09  & Cu${2}$–O11–Cu${2}$ = 94.5  & $J_{22}^{\rm b}$ &  K$_{2}$Cu$_{3}$(MoO$_{4}$)$_{4}$: $\angle\mathrm{Cu\!-\!O\!-\!Cu} = 90.36^\circ$, $J$ = 4.176 K, $d_{\rm Cu\!-\!Cu}\sim3.11$ \AA \ \  \cite{PhysRevB.111.144420}. \\ \cline{1-3}
		Cu${3}$–Cu${3}$ = 3.05  & Cu${3}$–O12–Cu${3}$ = 95.9 & $J_{33}^{\rm a}$ & \\ \cline{1-3}
		Cu${3}$–Cu${3}$ = 3.08  & Cu${3}$–O12–Cu${3}$ = 94.4 & $J_{33}^{\rm b}$ & \\ \hline
	\end{tabular}
\end{table*}

\begin{figure*}
	\centering
	\includegraphics[width=\textwidth]{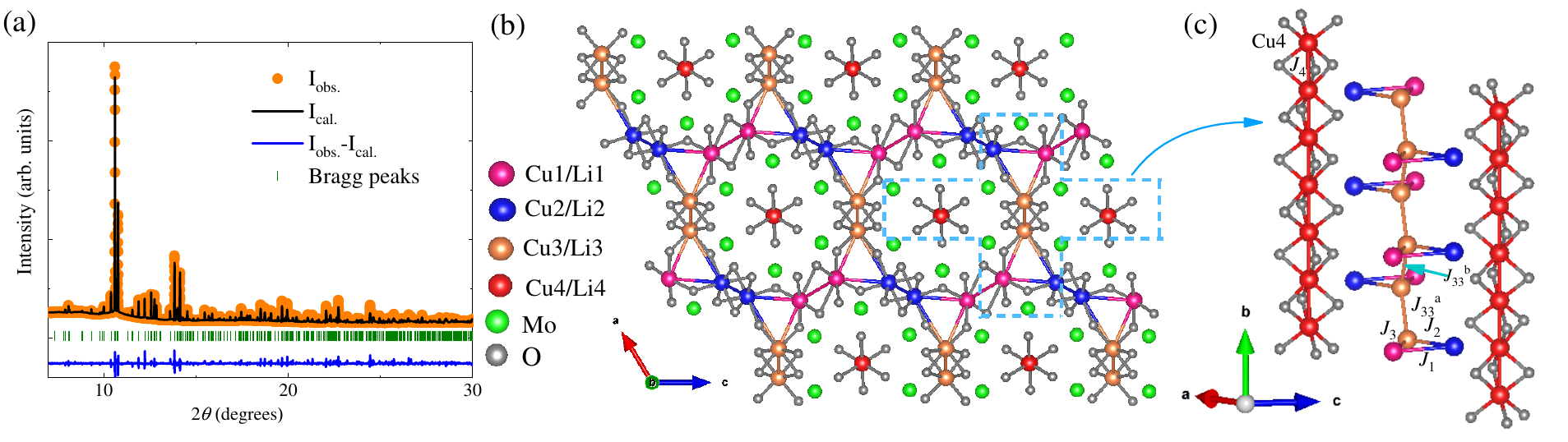}
	\caption{(a) Synchrotron XRD pattern at 120 K together with the Rietveld refinement. Experimental data are shown as filled circles, the calculated profile as a black line, and their difference (obs.–cal.) as a blue line. The Bragg peak positions are marked by olive vertical bars. (b) Schematic of the 3D version of a distorted stuffed hyper-star lattice of Cu$^{2+}$ ions projected along the $b$-axis. The Cu1, Cu2, and Cu3 sites are interconnected via O ions, forming superexchange pathways of the type Cu–O–Cu. In contrast, the Cu4 site resides at the center of the polygons and nearly isolated from the other Cu sites. (c) The dashed region in Fig.~\ref{STFig}(b) is shown with the $b$-axis perpendicular to the $ac$ plane, highlighting a distorted triangular spin network of Cu$^{2+}$ ions formed by the Cu1, Cu2, and Cu3 sites. These units are vertically connected by inter-triangular bonds, generating zigzag chains of Cu ions running along the $b$-axis. An example of a Cu3 chain is shown; similar chains are also formed by the Cu1 and Cu2 sites. In contrast, the Cu4 sites form a linear chain of Cu$^{2+}$ ions.
	 }{\label{STFig}}.
\end{figure*} 
\section{RESULTS}
\subsection{Rietveld refinement and crystal structure} \label{ST}
To confirm the phase purity and crystal structure of the polycrystalline samples of  LCMO, Rietveld refinement of the synchrotron XRD data was performed using the GSAS software~\cite{Toby:hw0089}. Figure~\ref{STFig}(a) presents the refinement profile, which confirms the absence of any detectable secondary phase. The initial atomic coordinates for the refinement were taken from the previous structural report~\cite{wiesmann1994crystal,Efremov1972DoubleMolybdates}, and the refined atomic positions are consistent with that report, as summarized in Table~\ref{SIF}. These results indicate that the present compound crystallizes in the monoclinic space group $P2_1/c$ with lattice parameters $a = 10.47$~\AA, $b = 5.01$~\AA, $c = 20.37$~\AA, $\alpha = \gamma = 90^\circ$, and $\beta = 120.94^\circ$.\\ 
The crystal structure viewed along the $b$-axis is shown in Fig.~\ref{STFig}(b). The Cu atoms occupy four inequivalent sites, highlighted in different colors, and partially share sites with Li$^{+}$ ions owing to the similar ionic radii of Li$^+$ and Cu$^{2+}$ ions. The refined occupancies indicate that the Cu$^{2+}$ ions occupy 30\%, 46\%, 49\%, and 80\% of the Cu${1}$, Cu${2}$, Cu${3}$, and Cu${4}$ sites, respectively (Table \ref{SIF}). Such unavoidable site mixing has also been observed  in related RS candidate Li$_4$CuTeO$_6$ \cite{Khatua2022}.  Minor deviations from nominal stoichiometry in the refined Cu/Li occupancies are attributed to the high scattering contrast between Li and the heavier Mo/Cu ions, which places these values within the uncertainty of the synchrotron x-ray refinement. Among the Cu sites, Cu2, Cu3, and Cu4 are coordinated by CuO$_6$ octahedra, whereas the Cu1 site forms a distorted CuO$_5$ pyramid. The Cu1, Cu2, and Cu3 sites are  interconnected through bridging oxygen atoms via Cu–O–Cu superexchange pathways (Fig.~\ref{STFig}(b)). In contrast, the Cu4 sites are relatively isolated from the other three and primarily linked through Cu4-O-Cu4 superexchange path. \\ \\ 
Most importantly, the superexchange pathways among the  Cu1, Cu2, and Cu3 sites form a distorted 3D star lattice. In an ideal 2D star lattice, the network can be viewed as a honeycomb lattice in which each vertex is replaced by a triangle, and these triangular motifs are connected by dimer bonds branching from  each vertex (see Ref.~\cite{PhysRevLett.99.247203,PhysRevB.70.174454,PhysRevB.98.155108,PhysRevB.81.134418}).  For a perfect 2D star lattice, the dimer bond lies flat within the plane of the triangular motifs. In contrast, in our system the dimer bonds are tilted out of the plane (e.g., $J_{33}^{a}$ and $J_{33}^{b}$ in Fig.~\ref{STFig}(c)), forming finite angles with the triangles. This deviation converts the planar dimer coupling into a zigzag chain, thereby yielding a 3D hyper-star  lattice.\\ Figure~\ref{STFig}(c) illustrates  this spin network, where two triangular motifs are connected by a zigzag chain that mediates inter-triangle coupling. For clarity, only the Cu3 sites are shown; however, the Cu1 and Cu2 sites form similar zigzag chains running parallel to the $b$ axis. On the other hand, the centers of the polygons constituting the hyper-star  lattice are occupied by the Cu4 site, which forms an isolated chain of Cu4 ions, as depicted in Fig.~\ref{STFig}(c),  yielding a stuffed hyper-star  lattice.\\
The resulting spin topology of the title compound can thus be regarded as a combination of two subsystems: a distorted hyper-star  lattice and a uniform Cu4 spin chain located at the centers of the hyper-star  lattice polygons. The bond angles and bond lengths for these sublattices are summarized in Table~\ref{tableangle}. Approximate exchange interactions are inferred from structurally analogous compounds reported previously (fourth column of Table~\ref{tableangle}).\\ Although calculating the exchange interactions in such systems is quite challenging due to site mixing, the Goodenough–Kanamori–Anderson rules together with the comparison with similar bonding geometries reveal that the intra-triangular interactions $J_{1}$, $J_{2}$, and $J_{3}$ (Fig.~\ref{STFig}(c)) with an order of magnitude of 60-90 K are significantly larger than those of the remaining exchange couplings. On the other hand, the inter-triangular interactions $J_{22}$ and $J_{33}$ are weak ferromagnetic (Fig.~\ref{STFig}(c)). In the absence of Li/Cu site mixing, our system may host a trimerized valence-bond crystal~\cite{PhysRevB.97.075146}. On the other hand, the Cu${4}$ chain is expected to host only weak interactions, arising primarily from a competition between direct exchange across the shortest Cu${4}$-Cu${4}$ separation and Cu${4}$-O-Cu${4}$ superexchange, while the remaining couplings are likely mediated only through superexchange pathways. Thus, the Cu4 subsystem may constitute weakly-coupled ferromagnetic chains.
\begin{figure*}[t]
	\centering
	\includegraphics[width=\textwidth]{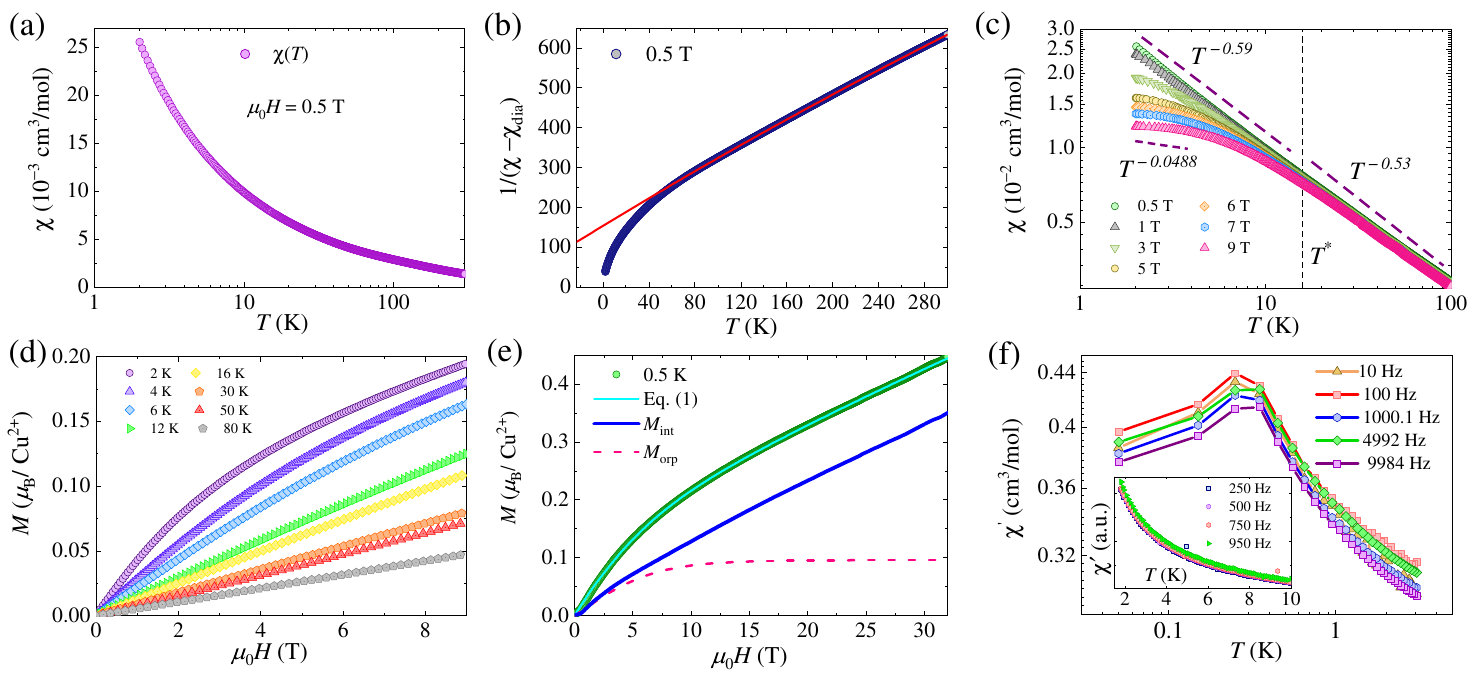}
	\caption{(a) Temperature-dependent magnetic susceptibility $\chi(T)$ of Li$_{2}$Cu$_{2}$(MoO$_{4}$)$_{3}$ at $\mu_{0}H = 0.5$ T.  (b) Temperature dependence of the inverse magnetic susceptibility after subtracting the diamagnetic contribution. The high-temperature Curie–Weiss fit is shown by a red  line.  (c) Log–log plot of $\chi(T)$  at various magnetic fields. The dashed vertical line (black) indicates the position of $T^{*} = 15.8$ K.  Above and below $T^{*}$, $\chi(T)$ follows a power-law behavior, as indicated by the dashed lines (purple). (d) Isothermal magnetization at various temperatures. (e) High-field magnetization data decomposed into intrinsic magnetization $M_{\mathrm{int}}$ and orphan-spin contribution $M_{\mathrm{orp}}$ using Eq.~(\ref{eq1}), as described in the text. (f) Real part of the \textit{ac} magnetic susceptibility at various frequencies in the low-temperature range. The inset shows the data above 2 K for lower frequencies.  }{\label{MAGFig}}.
\end{figure*} 
\subsection{Magnetic susceptibility}
Figure~\ref{MAGFig}(a) shows the temperature dependence of the magnetic susceptibility $\chi(T)$, measured at $\mu_{0}H = 0.5$ T down to 2 K. Upon lowering the temperature,  $\chi{(T)}$ increases monotonically, without any clear signature of long-range magnetic ordering. The low-temperature upturn is attributed to contributions from weakly coupled spins or orphan spins arising from magnetic-site vacancies in LCMO~\cite{Vasiliev2018}. This interpretation is further supported by the non-linear behavior of the isothermal magnetization at low temperatures, as discussed below.\\
To determine the dominant magnetic exchange interactions, the temperature dependence  of the inverse magnetic susceptibility $1/\chi(T)$ is plotted in Fig.~\ref{MAGFig}(b) after subtracting the total diamagnetic contribution from all ions, $\chi_{\rm dia} = -2.3 \times 10^{-3}$ cm$^{3}$/mol \cite{Bain2008}. The high-temperature linear region of $1/\chi(T)$ follows a Curie–Weiss (CW) law, $\chi = \chi_{\rm VV} + C/(T-\theta_{\rm CW})$, where $\chi_{\rm VV} = 1.78 \times 10^{-4}$ cm$^{3}$/mol is the temperature-independent Van Vleck susceptibility, $C = 0.54$ cm$^{3}$K/mol is the Curie constant, and $\theta_{\rm CW} = -85.76$ K is the CW temperature. The large negative value of $\theta_{\rm CW}$ indicates dominant antiferromagnetic interactions in LCMO. The effective magnetic moment was estimated as $\mu_{\rm eff} = \sqrt{8C} = 2.07 \  \mu_{\rm B}$, corresponding to $s$ = 1/2 moment with a Landé $g$ factor of 2.39.
Below 80 K, the experimental data deviate from the CW behavior, implying the development of antiferromagnetic spin correlations, as further supported by the magnetic specific heat discussed in Sec.~\ref{HC}.\\
As discussed in Sec.~\ref{ST}, the system comprises two distinct spin topologies. 
To clarify which topology, either the 3D hyper-star  lattice or the Cu$4$ spin chain, dominates the antiferromagnetic interactions, we performed DFT calculations accounting for possible Cu/Li site mixing. 
In these calculations, Cu$4$ and Cu$1$ sites were assumed to be fully occupied by Cu$^{2+}$ ions, while Li$2$ and Li$3$ sites were occupied by Li$^{+}$ ions to maintain charge neutrality according to the stoichiometric formula unit. 
Using this configuration, DFT calculations with an on-site Hubbard parameter of \(U = 6~\mathrm{eV}\) yield an exchange interaction strength of \(J_{4} \approx 1.28~\mathrm{K}\) (Cu$4$--Cu$4$), corresponding to an approximate CW temperature of \(\theta^{\rm chain}_{\rm CW} \approx -0.64~\mathrm{K}\) within the mean-field approximation. However, the experimentally observed large negative \(\theta_{\rm CW}\) = $-85.76$ K suggests that the dominant antiferromagnetic exchange originates from the hyper-star  lattice spin topology rather than from the spin-chain network.
 The exchange interactions within the hyper-star  lattice were not explicitly calculated, as site mixing and the associated charge-balance constraints among the Cu$1$, Cu$2$, and Cu$3$ sites introduce considerable complexity. Notwithstanding, we recall that the qualitatively deduced exchange interactions of $J_{1}$, $J_{2}$, $J_{3}$ $\sim$ 60-90 K match well with the CW temperature.\\ Figure~\ref{MAGFig}(c) displays the temperature dependence of $\chi(T)$ at various magnetic fields below 100 K. Upon cooling below $|\theta_{\rm CW}|$, $\chi(T)$ follows a $T^{-0.53}$ power-law behavior down to $T^{*} = 15.8 $ K with weak field dependence. This sub-Curie $\chi(T)$ increase alludes to the presence of random exchange interactions. At lower temperatures, a steeper power-law increase of $T^{-0.59}$ at 0.5 T signals enhanced RS-like behavior possibly due to the increasingly enhanced contribution of weakly interacting spins, resulting in a broader distribution of exchange interactions. While $\chi(T)$ above $T^{*}$ remains nearly field independent, below $T^{*}$ it shows strong field dependence, with the exponent evolving from 0.59 at 0.5 T to 0.048 at 9 T, implying a progressive quenching of weakly coupled RS with increasing magnetic fields.\\ 
To further confirm the existence of randomly interacting spins in LCMO, the isothermal magnetization $M(H)$ was measured at several temperatures, as shown in Fig.~\ref{MAGFig}(d). Above $T^{*}$, $M(H)$ displays a linear field dependence, consistent with $T<$ $|\theta_{\rm CW}|$. In contrast, below $T^{*}$, the magnetization deviates from linearity without exhibiting any hysteresis, witnessing the subdominant yet finite contribution of  orphan spins. Throughout this paper, orphan spins in the RS context collectively refer to weakly coupled paramagnetic or defect spins and vacancy-induced quasispins. \\ \begin{figure*}
	\centering
	\includegraphics[width= 17.5 cm, height=4.5 cm]{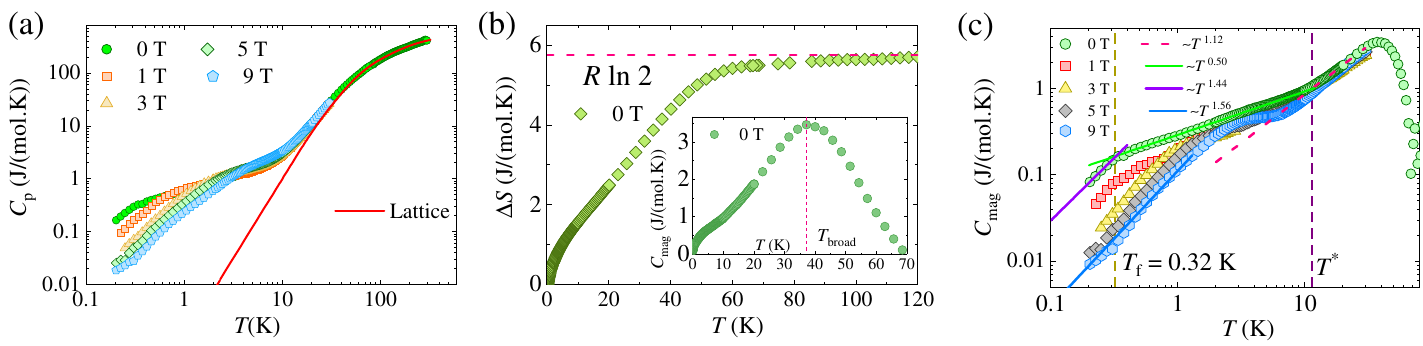}
	\caption{(a) Temperature dependence of the specific heat at several magnetic fields, where the solid red line corresponds to  the lattice contribution. (b) Magnetic entropy as a function of temperature at zero field.  The dashed pink horizontal line indicates the expected entropy $R\ln2$ for an $s=1/2$ system. The inset shows the magnetic specific heat as a function of temperature. (c) Log–log plot of the magnetic specific heat as a function of temperature at various magnetic fields. In zero field, three distinct power-law regimes below the broad maximum are identified, as indicated in the legend, and separated by dashed vertical lines at $T^{*}=11.11$ K and $T_{\rm f}=0.32$ K.   }{\label{HCFig}}.
\end{figure*} 
High-field magnetization measurements up to 35~T were conducted to examine possible field-induced correlated states in this  magnetic site-diluted 3D system~\cite{PhysRevB.97.075146}. As illustrated in Fig.~\ref{MAGFig}(e), the magnetization reaches $0.45$ $ \mu_{\rm B}$ at 31 T, whose extrapolation yields the expected full saturation value of $1$  $\mu_{\rm B}$ at fields close to 85 T, in accord with $|\theta_{\rm CW}|$. Although ideal 2D star lattices are predicted to host field-induced correlated states, such features are absent in the present system, likely due to the interplay of 3D exchange interactions and intrinsic magnetic site disorder~\cite{PhysRevB.97.075146}. In line with the low-field observations below $T^{*}$ (Fig.~\ref{MAGFig}(d)), the non-linear magnetization $M(H)$ at 0.5 K (Fig.~\ref{MAGFig}(e)) is governed by orphan spins and
its concentration can be determined by using the following phenomenological model~\cite{PhysRevLett.109.117203,PhysRevB.50.16754}
\begin{equation}
M(H) = \frac{\eta_{\rm orp} g \mu_{B}}{2} 
\tanh \left[ \frac{g \mu_{B} \mu_{0} H}{k_{B}(T - \theta^{*}_{\rm orp})} \right] 
+ A (\mu_{0} H)^{n},
\label{eq1}
\end{equation} 
where the first term represents a modified Brillouin function describing the magnetization of weakly interacting orphan spins, while the second term captures the power-law field dependence arising from random exchange interactions~\cite{PhysRevB.22.1305,PhysRevB.50.16754,PhysRevLett.43.1434}.  In Eq.~(\ref{eq1}), $\eta_{\rm orp}$ refers to the number of orphan spins, $\mu_{\rm B}$ is the Bohr magneton, $g$ denotes the Landé $g$-factor, and $\theta^{*}_{\rm orp}$ represents  the effective interaction energy scale between orphan spins, while $A$ and $n$ are  fitting parameters. The solid cyan line in Fig.~\ref{MAGFig}(e) represents the fits obtained using Eq.~(\ref{eq1}) assuming $g = 2$, and the corresponding parameters yield $\eta_{\rm orp} = 9.63(1)\,\%$, $\theta^{*}_{\rm orp} = -8.52(1)\,\mathrm{K}$, $n = 0.853(2)$, and $A = 0.0181(2)$. The obtained exponent $n$ = 0.853, which is smaller than 1, reflects the contribution arising from randomness in the exchange interactions \cite{PhysRevB.22.1305}, and may be indicative of RS–like correlations.\\
In systems with quenched disorder, spin freezing often emerges at low temperatures. To inspect whether such freezing occurs in the present case, we performed frequency-dependent \textit{ac} magnetic susceptibility measurements over a wide temperature range. The data between 2 and 10 K are shown in the inset of Fig.~\ref{MAGFig}(f). The absence of any frequency dependence rules out spin freezing down to 2 K. In contrast, the low-temperature \textit{ac} magnetic susceptibility data collected between 0.05 K and 3.4 K  (Fig.~\ref{MAGFig}(f)) reveal a cusp around $T_{\rm f} \approx 0.3$ K. With increasing frequency, this cusp shifts to higher temperatures, indicating that partial spin freezing takes place.
\subsection{Specific heat}\label{HC}
In order to confirm the absence of long-range magnetic order and probe low-energy magnetic excitations across different energy scales, specific-heat measurements were carried out down to 0.2~K. Figure~\ref{HCFig}(a) presents the temperature dependence of the specific heat  at several magnetic fields. The absence of a $\lambda$-type anomaly rules out any thermodynamic phase transition in LCMO down to the lowest measured temperature. \\
In general, the specific heat $C_{p}(T)$ in magnetic insulators comprises three contributions, namely the magnetic, lattice, and  Schottky terms relevant under finite fields.  Given the unavailability of a nonmagnetic analogue of LCMO, the lattice contribution was determined by fitting the data (solid red line in Fig.~\ref{HCFig}(a)) with a model comprising one Debye and three Einstein terms i.e.,
\begin{equation}
\begin{split}
C_{\rm latt}(T) &= C_D \left[ 9R \left(\frac{T}{\theta_D} \right)^3 
\int_0^{\theta_D/T} \frac{x^4 e^x}{(e^x - 1)^2} \, dx \right] \\
&\quad + \sum_{i=1}^{3} C_{E_i} \left[ R \left( \frac{\theta_{E_i}}{T} \right)^2 
\frac{e^{\theta_{E_i}/T}}{(e^{\theta_{E_i}/T} - 1)^2} \right],
\end{split}
\end{equation}

where $\theta_{D} = 180 \pm 6$~K is the Debye temperature, and $\theta_{E_1} = 282 \pm 2$~K, $\theta_{E_2} = 293 \pm 4$~K, and $\theta_{E_3} = 751 \pm 8$~K represent the Einstein temperatures of three optical phonon modes, with $R$ being the molar gas constant. To minimize the number of free parameters, $C_D$ was set to 3, corresponding to the three acoustic phonon modes. The best fit is obtained with $C_{E_1} = 4$, $C_{E_2} = 5$, and $C_{E_3} = 10$, which together account for the 54 optical phonon modes expected from the $(3n-3)$ branches for $n=19$ atoms in LCMO~\cite{PhysRevB.110.184402}.\\ After subtracting the lattice contribution, the resulting magnetic specific heat $C_{\rm mag}(T)$ is presented in the inset of Fig.~\ref{HCFig}(b). $C_{\rm mag}(T)$ increases below 70 K and displays a broad maximum near 40 K, suggesting the presence of short-range spin correlations arising from antiferromagnetic Heisenberg interactions in the the hyper-star lattice spin network~\cite{PhysRevB.109.125120}. We note that a deviation from the CW fit emerges below 80 K, which coincides with the growth of $C_{\rm mag}(T)$ below 70 K. The change in magnetic entropy was determined by integrating $C_{\rm mag}(T)/T$, as shown in Fig.~\ref{HCFig}(b). The total entropy recovered near the CW temperature is $\sim$5.67 J/(mol K), in good agreement with the expected $R\ln(2S+1)$ for an $s = 1/2$ system.\\
Figure~\ref{HCFig}(c) depicts the temperature dependence of $C_{\rm mag}(T)$ at various magnetic fields after subtracting the  Schottky contribution (see Appendix A). The coexistence of distinct energy scales associated with the two sublattices  is particularly evident in the zero-field data. For instance, below the broad maximum, $C_{\rm mag}(T)$ follows an almost linear $T^{1.12}$ dependence, terminating near $T^{*} \approx 11.1$ K, indicative of low-energy gapless excitations  that persist down to $T^{*}$.
 At lower temperatures, $C_{\rm mag}(T)$ evolves into a $T^{0.50}$ dependence over the range 0.32–11.1 K, pointing to the prevailing contributions of a RS–like state, consistent with $\chi(T)\sim T^{-0.59}$ behaviour below $T^{*}$. Upon further cooling ($T < 0.32$ K), $C_{\rm mag}(T)$ undergoes a curvature change, approaching a $T^{1.44}$ dependence. This crossover coincides with the spin-freezing transition observed in the \textit{ac} magnetic susceptibility at $T_{\rm f} \approx 0.32$ K, below which quasi-frozen states manifest through a sub-quadratic temperature dependence. As the magnetic field increases, the subtle low-temperature slope below 0.32 K weakens progressively, indicating the suppression of the quasi-frozen state while retaining the sub-quadratic temperature dependence characteristic of exotic gapless excitations, which persists even up to 9 T.\\ \\
Note that the characteristic temperature $T^{*}$ identified from $C_{\rm mag}(T)$ is slightly lower than that observed in $\chi(T)$ and other microscopic experiments ($T^{*} = 15.8~\mathrm{K}$) discussed below. This difference likely arises from the varying sensitivities of the experimental probes. However, we consider $T^{*} = 15.8~\mathrm{K}$ as the reference value derived from microscopic measurements.\begin{figure}[t]
	\centering
	\includegraphics[width=0.5\textwidth]{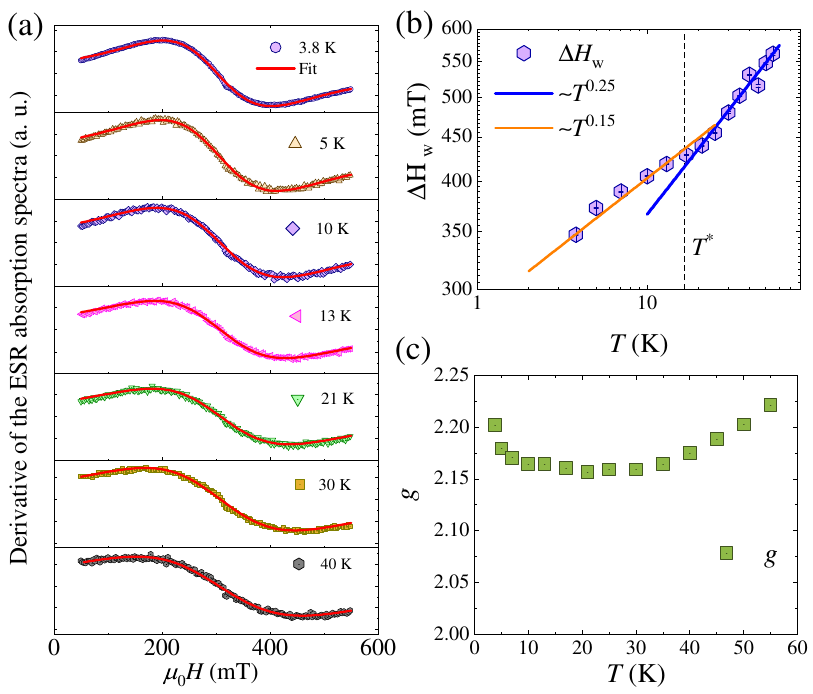}
	\caption{(a)Derivative of the ESR absorption spectra of Li$_{2}$Cu$_{2}$(MoO$_{4}$)$_{3}$ at selected temperatures where the Lorentzian line-shape fitting is shown by the solid red lines.  (b) Temperature dependence of the ESR linewidth $\Delta H_{w}$ that shows a crossover in its power-law dependence across $T^{*}$ = 15.8 K (dashed vertical line). (c) Temperature dependence of the $g$ factor, exhibiting a non-monotonic variation. }{\label{ESRFig}}.
\end{figure} \begin{figure*}[t]
\centering
\includegraphics[width=\textwidth]{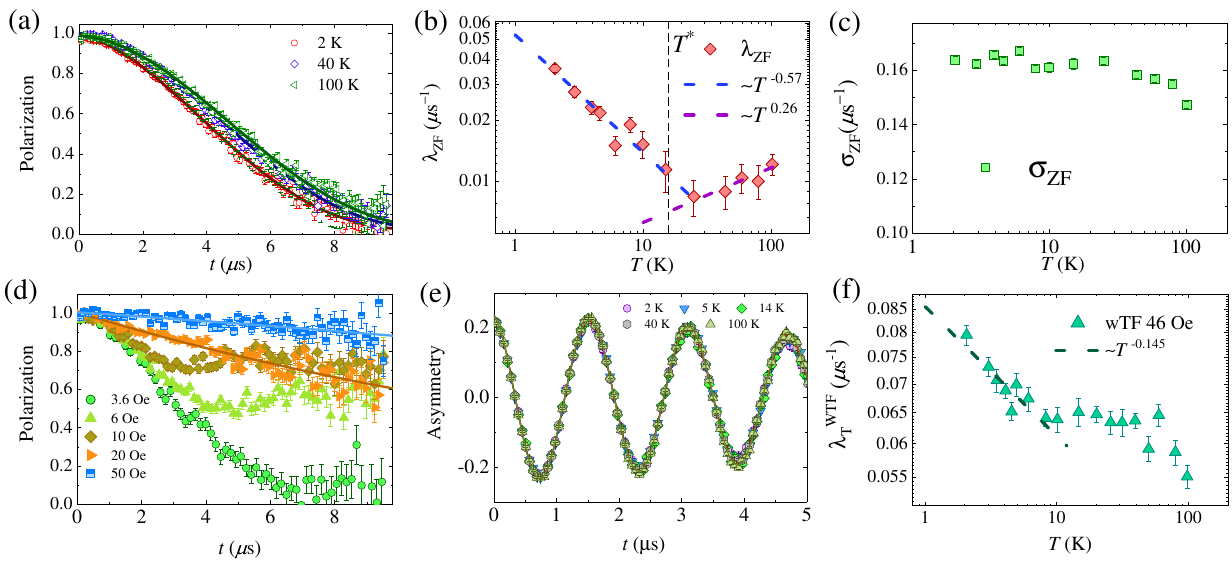}
\caption{(a) Time evolution of normalized  muon spin polarization  in zero-field at three temperatures. (b) and (c) Temperature dependence of the muon spin relaxation rate of electronic origin ($\lambda_{\rm ZF}$) and nuclear origin ($\sigma_{\rm ZF}$). The dashed lines indicate the crossover in spin–spin correlations across $T^{*}$ = 15.8 K, highlighting distinct power-law regimes as discussed in the text. (d) Time evolution of normalized  muon spin polarization  in several longitudinal fields at 2 K.  (e) Time dependence of muon spin asymmetry at a weak-transverse  field $H_{\rm TF}$ = 46 Oe  at several temperatures. (f) Temperature dependence of the transverse relaxation rate extracted from weak transverse-field measurements.  }{\label{LCMOFig5}}.
\end{figure*} 
\subsection{Electron spin resonance}
To further elucidate the thermal evolution of spin–spin correlations, we performed X-band ESR measurements down to 3.8 K. As shown in Fig.~\ref{ESRFig}(a), the ESR spectra recorded at various temperatures are well reproduced by  a derivative of a Lorentzian curve. The ESR spectra exhibit pronounced broadening at high temperatures, followed by gradual narrowing as the temperature decreases. For $T>$ 60 K, the linewidth becomes too broad to be resolved within the X-band frequency window.\\ \\ As the ESR linewidth reflects the combined effects of spin–spin and spin–lattice relaxation, its temperature evolution, as captured by the ESR linewidth $\Delta H_{w}(T)$, provides valuable information on the underlying spin dynamics, as illustrated in Fig.~\ref{ESRFig}(b).  Upon lowering the temperature below 60 K, $\Delta H_{w}$ decreases gradually, following a power-law dependence, 
$\Delta H_{w} \propto T^{0.25}$, down to $T^{*}$ = 15.8 K. For $T \gg J$, the ESR linewidth is usually expected to remain temperature independent. However, the observed anomalous temperature dependence indicates the influence of spin–spin correlations and spin–lattice coupling.   Furthermore, in transition--metal--based compounds, the spin--lattice relaxation 
contribution to the ESR linewidth typically leads to either an increase with temperature 
(via direct or Raman processes) or a saturation at low temperatures~\cite{abragam2012electron}. The sublinear narrowing observed here cannot be accounted  by conventional 
spin--lattice relaxation.  Instead, it points to the dominant role of spin--spin relaxation 
processes, possibly associated with static short-range order in low-dimensional magnets~\cite{PhysRevB.16.3036}.  \\ \\Below $T^{*}$, the linewidth exhibits a further decrease, following a different 
power-law trend of $\sim T^{0.15}$, indicating the onset of an another correlated 
regime. Unlike the inhomogeneous broadening commonly observed in RS candidates~\cite{PhysRevB.107.214404}, the present system exhibits a two-stage power-law narrowing of the ESR linewidth upon cooling. A similar two-step linewidth evolution has been reported for the alternating kagome bilayer compound Ca$_{10}$Cr$_{7}$O$_{28}$ \cite{PhysRevB.111.125144}, where it was attributed to the emergence of two magnetic sublattices with distinct correlation strengths. In the gigahertz frequency window, the spin dynamics appears to be dominated by exotic spin correlations, rather than inhomogeneous RS spin dynamics.\\ 
Figure~\ref{ESRFig}(c) shows a weakly non-monotonic temperature dependence of the \( g \)-factor (\(\approx 2.15\!-\!2.20\)), exhibiting a shallow minimum around $T^{*}$ and slight upturns at both lower and higher temperatures. Notably, \(C_{\rm mag}(T)\) changes its power-law behavior around $T^{*}$, closely aligning with the \(g\)-factor minimum. Throughout the measured temperature range, the effective \(g\) value remains typical of Cu\(^{2+}\)-based systems with a quenched orbital moment~\cite{PhysRevMaterials.8.094404}. In exchange-coupled insulating magnets, the variation of \( g \) on both sides of the minimum indicates a crossover in spin–spin correlations, consistent with the changes observed in the linewidth and \( C_{\rm mag}(T) \) around $T^{*}$.\\
Altogether, the intra-triangular spin correlations progressively evolve into a more RS-like state that coexists with spin-chain correlations as weak residual interactions become relevant at lower temperatures.
\subsection{Muon spin relaxation}
To further investigate the low-energy spin dynamics associated with magnetic site dilution, 
$\mu$SR measurements were performed down to 2~K. 
$\mu$SR is a highly sensitive local probe, in which fully polarized positive muons are implanted into the sample 
to monitor the microscopic spin dynamics through the time evolution of the muon spin polarization~\cite{yaouanc2011muon}.\\
Figure~\ref{LCMOFig5}(a) depicts the time evolution of the normalized muon spin polarization at various temperatures in ZF. A slow relaxation is observed between 2 K and 120 K; however, the absence of coherent oscillations indicates the lack of long-range magnetic order  at least down to 2 K. Moreover, no indication of a ``1/3’' tail in the ZF asymmetry further rules out any conventional spin freezing for $T \geq 2$ K, consistent with thermodynamic results~\cite{PhysRevB.31.546}.\\
The ZF spectra are well modeled by the following polarization function (solid lines in Fig.~\ref{LCMOFig5}(a))
\[
P (t) = \exp\!\left[-\tfrac{1}{2}(\sigma_{\rm ZF} t)^{2}\right]\exp(-\lambda_{\rm ZF} t),
\]
where the Gaussian term accounts for the static nuclear dipolar fields characterized by the relaxation rate $\sigma_{\rm ZF}$, while the exponential term represents the slow relaxation of electronic origin with the muon spin relaxation rate $\lambda_{\rm ZF}$. A comparable  muon spin polarization, arising from the combined effects of nuclear dipolar fields and electronic spins, has been observed in the Li/Cu site-mixing RS candidate Li$_4$CuTeO$_6$~\cite{Khatua2022}. \\ 
The estimated $\lambda_{\rm ZF}$ is shown in Fig.~\ref{LCMOFig5}(b) as a function of temperature. In most 3$d$ transition-metal compounds, the ZF muon spin relaxation rate of electronic origin provides a measure of spin–spin correlations. In LCMO, above $T^{*}$ = 15.8 K, $\lambda_{\rm ZF}$ exhibits only a weak temperature dependence, following a $T^{0.26}$ power-law behavior, which signals the persistence of a characteristic spin–spin correlations at elevated temperatures. Consistently, the ESR linewidth also follows a $T^{0.25}$ dependence above $T^{*}$. Such weak power-law behavior ($\lambda_{\rm ZF} \sim T^{0.26}$ and ESR linewidth $\sim T^{0.25}$) is often reported in 1D Luttinger liquids and 2D frustrated magnets hosting specific spinons~\cite{PhysRevLett.119.137205} and in LCMO this correlated regime extends
down to  $T^{*}$.\\
Upon further cooling below $T^{*}$, $\lambda_{\rm ZF}$ begins to increase and follows a distinct power-law behavior, $\lambda_{\rm ZF} \sim T^{-0.57}$. Notably, this growth of $\lambda_{\rm ZF}(T)$  is opposite to the ESR linewidth narrowing, highlighting the frequency-dependent spin dynamics. In the MHz time window, the influence of quenched disorder and associated localized orphan spin becomes more prominent~\cite{PhysRevB.106.L140202}. These vacancy-induced orphan spins most likely originating from Li/Cu site-mixing in LCMO, interact with the surrounding bulk spins in a highly inhomogeneous manner. As such, slow spin dynamics dominated by vacancy-induced orphan spins is responsible for the low-temperature power-law enhancement of 
$\lambda_{\rm ZF}$~\cite{shimokawa202,PhysRevLett.126.037201}. \\
Figure~\ref{LCMOFig5}(c) shows the temperature dependence of the Gaussian relaxation rate, which remains nearly temperature-independent over the measured temperatures, reflecting the maximum width of the static nuclear field distribution of about $\sim 2$ Oe.  To confirm that this Gaussian relaxation originates from nuclear dipolar fields, we performed LF-$\mu$SR measurements at 2 K, as shown in Fig.~\ref{LCMOFig5}(d). Upon increasing the applied LF beyond the static nuclear field ($\gg 2$ Oe), the muon spin relaxation  progressively decouples from the static nuclear fields. At around 20 Oe ($\approx$ 10 times of nuclear field), a nearly complete decoupling is achieved, leaving  only a weak exponential relaxation component of  electronic origin. The solid lines in Fig.~\ref{LCMOFig5}(d) represent fits using the simple exponential polarization function $P(t)_{\rm LF} = \exp(-\lambda_{\rm LF} t)$, which yield the relaxation rates of $\lambda_{\rm LF} = 0.013~\mu\text{s}^{-1}$ at 20 Oe and $0.0023~\mu\text{s}^{-1}$ at 50 Oe. Notably, even under fields exceeding the static nuclear field by more than an order of magnitude, a finite muon spin relaxation persists, evidencing the presence of slow electronic spin fluctuations at low temperatures.\\
To further investigate the low-temperature spin dynamics, we carried out wTF-$\mu$SR measurements at several temperatures. Figure~\ref{LCMOFig5}(e) presents the time evolution of the muon spin asymmetry in an external field of $H_{\rm TF} = 46$ Oe, applied perpendicular to the initial direction of muon spin polarization. In this geometry, the muon spins precess coherently around the transverse  field $H_{\rm TF}$.
 As the applied field is much larger than the width of the static nuclear field obtained from ZF, the observed relaxation beyond the nuclear contribution must stem from paramagnetic spins, in line with the ZF data that show no evidence of static electronic fields. As evident from Fig.~\ref{LCMOFig5}(e), the full asymmetry ($A_{0}$) persists down to 2 K, confirming the lack of static electronic fields in agreement with the ZF data. To analyze the wTF spectra, we employed a damped oscillatory function,
 \begin{equation}
 A_{\rm wTF}(t) = A_{0}  \cos(\gamma_{\mu} H_{\rm TF} t + \phi) e^{-\lambda_{\rm TF} t},
 \end{equation}
 in which the muon spins precess at frequency $\omega = \gamma_{\mu} H_{\rm TF}$, $\phi$ represents the  phase shift, and $\lambda_{\rm TF}$ quantifies the relaxation rate arising from slow spin fluctuations of electronic origin. \\
 \begin{figure*}
 	\centering
 	\includegraphics[width= 17.5 cm, height=4.5 cm]{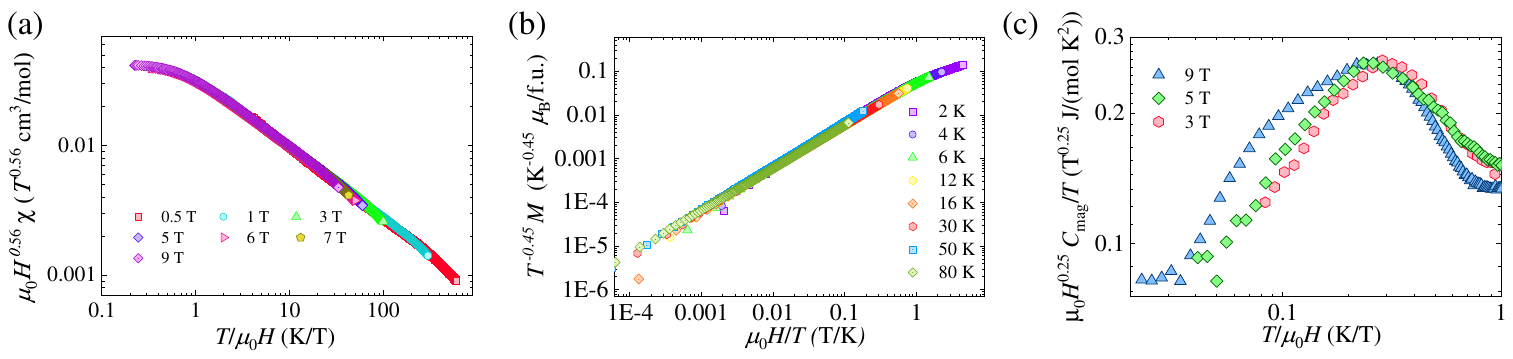}
 	\caption{(a) Magnetic susceptibility $\chi(T)$ rescaled by magnetic field $\mu_{0}H^{0.56}$ and plotted against $T/\mu_{0}H$, showing data collapse onto a universal-like curve. (b) Isothermal magnetization $M$ multiplied by $T^{-0.45}$ as a function of $\mu_{0}H/T$, demonstrating consistent scaling over a wide temperature range. (c) Scaled magnetic specific heat $\mu_{0}H^{0.25}C_{\mathrm{mag}}/T$ versus $T/\mu_{0}H$ at selected fields, highlighting collapse onto a common trend. All panels are presented in a log–log scale. }{\label{DC}}.
 \end{figure*}
 The temperature dependence of the transverse relaxation rate $\lambda_{\rm TF}$ is shown in Fig.~\ref{LCMOFig5}(f). For $T > T^{*} $, $\lambda_{\rm TF}$ remains nearly temperature-independent, reflecting the response of randomly oriented paramagnetic spins. In contrast, a pronounced enhancement  appears between 2 and 10 K, indicating the presence of a crossover regime that may be related to the correlated orphan spins behavior inferred from the ZF measurements. A direct comparison reveals that the ZF  relaxation rate $\lambda_{\rm ZF}$ (Fig.~\ref{LCMOFig5}(b)) is  weaker than the transverse field relaxation rate $\lambda_{\rm TF}$  (Fig.~\ref{LCMOFig5}(f)). This is related to the fact that while the ZF relaxation reflects only dynamic field fluctuations, the transverse relaxation is additionally affected by inhomogeneous local fields induced by the applied transverse field. Furthermore, the overall temperature dependence of $\lambda_{\rm TF}$ ($\sim T^{-0.14}$) is weaker than that of $\lambda_{\rm ZF}$ ($\sim T^{-0.57}$), possibly due to partial quenching of weakly coupled spin dynamics by the applied transverse field.
 \begin{table}[b]
 	\caption{Temperature-dependent power-law behaviour for various physical quantities in Li$_2$Cu$_2$(MoO$_4$)$_3$.}
 	\centering
 	\small
 	\setlength{\tabcolsep}{3.8pt} % reduce horizontal spacing
 	\renewcommand{\arraystretch}{1.6} % reduce vertical spacing
 	\begin{tabular}{|p{1.9cm}|p{2.5cm}|p{2.7cm}|}
 		\hline
 		\centering\textbf{Physical}\\[-2pt]\centering\textbf{Quantity} &
 		\centering\textbf{Intra-triangle}\\[-2pt]\centering(\textit{$T > T^{*}$}) &
 		\centering\textbf{Random-singlet}\\[-2pt]\centering( \textit{$T_{\rm f} < T < T^{*}$}) \tabularnewline
 		\hline
 		\centering$\chi(T)$ & \centering$T^{-0.53}$ & \centering$T^{-0.59}$ \tabularnewline
 		\hline
 		\centering$C_{\rm mag}(T)/T$ & \centering$T^{0.12}$ & \centering$T^{-0.50}$ \tabularnewline
 		\hline
 		\centering$\lambda_{\rm ZF}(T)$ (MHz) & \centering$T^{0.26}$ & \centering$T^{-0.57}$ \tabularnewline
 		\hline
 		\centering$\Delta H_{w}(T)$ (GHz) & \centering$T^{0.25}$ & \centering$T^{0.15}$ \tabularnewline
 		\hline
 		\multicolumn{3}{|c|}{\textbf{Quasi-frozen regime:} $T < T_{\rm f}$, $C_{\rm mag}(T)/T \propto T^{0.44}$} \tabularnewline
 		\hline
 	\end{tabular}
 	\label{tab:exponents}
 \end{table}
\section{Discussion}
Vacancy-induced quenched disorder presents fundamental challenges in understanding correlated quantum states. In the present compound, Li/Cu site mixing introduces intrinsic disorder that perturbs the exchange network of the clean hyper-star lattice while largely preserving its frustrated character.\\  It is noteworthy that despite the substantial Li/Cu site mixing, the system does not appear to reside at a magnetic percolation threshold. Instead, the Cu$^{2+}$  moments form an extended, randomly connected 3D network that effectively fragments into low-dimensional magnetic segments. In this geometry, the primary source of frustration associated with the ideal triangular plaquettes is mostly suppressed. Nevertheless, residual frustration may persist due to bond disorder-induced random exchange interactions, which are well-established as a driver of spin frustration in 1D RS  systems \cite{PhysRevB.35.419}.\\
 At high temperatures, dominant antiferromagnetic correlations originate from strong yet diluted intra-triangle exchange interactions. Upon cooling, a RS–like  state develops through a distribution of inter-triangle exchange couplings and weakly interacting Cu4 spin chains. Ultimately, the spins undergo progressive freezing driven by the residual 3D exchange interactions linking the hyper-star lattice and the Cu4 chain network.
\\
The dominance of antiferromagnetic intra-triangle interactions and the emergence of orphan spins driven RS–like state are  manifested in the $\chi(T)$ data. In the temperature range 15–100 K, $\chi(T)$ remains nearly field independent and follows a power-law behavior $\chi(T) \sim T^{-0.53}$, consistent with random magnetism dominated by the magnetic segment coupled by intra-triangle antiferromagnetic interactions~\cite{PhysRevResearch.6.023225}. At lower temperatures, however, the growing contribution from weak residual interactions enhances the magnetic susceptibility through a distribution of exchange interactions, as reflected by $\chi(T) \sim T^{-0.59}$ below $T^{*}$. With increasing magnetic field, the low-temperature susceptibility is drastically modified, as reflected by the change of the exponent from 0.59 to 0.0488, implying a gradual polarization of the orphan spins under applied fields below $T^{*}$. The spin dynamics arising from low-dimensionally segmented antiferromagnetic interactions at high temperatures are further corroborated by the broad maximum in $C_{\rm mag}(T)$~\cite{PhysRevLett.134.226701}, along with the power-law behavior of the ESR linewidth and the zero-field muon spin relaxation rate above $T^{*}$, indicating the development of short-range spin correlations typical of low-dimensional Heisenberg antiferromagnets~\cite{PhysRevLett.134.226701}.\\ \\
Before discussing further evidence for an intermediate RS–like state below $T^{*}$, it is worth noting that, despite the presence of both vacancy and bond disorder, we do not observe a clear $\chi(T) \propto C/T$ Curie-tail at low temperatures. Such behavior is commonly seen in 1D systems, where it often originates from extraneous orphan spins \cite{Vasiliev2018}. In our case, the magnetization data indicate that orphan spins do exist, but they are an intrinsic feature of the site-diluted system rather than isolated paramagnetic impurities~\cite{PhysRevB.105.L060406,PhysRevB.106.L140202}. Instead of producing a simple $1/T$ divergence, these orphan spins are weakly correlated in the random-bond background, forming singlets of varying energy scales.  This redistribution of  exchange energy suppresses the Curie-like tail and drives the system toward a RS–like state, characterized by $\chi(T) \sim T^{-\alpha}$, $C_{\rm mag}(T) \sim T^{1-\alpha}$ and $\lambda_{\rm ZF}(T) \sim T^{-\alpha}$, where the low-energy density of states follows $D(E) \propto E^{-\alpha}$~\cite{PhysRevX.8.041040,PhysRevLett.126.037201,PhysRevB.111.014409,PhysRevLett.127.127201}.\\
In addition to the power-law behavior of $\chi(T)$, the magnetic specific heat below $T^{*}$ follows $C_{\rm mag}(T)/T \sim T^{-0.5}$ down to $T_{\rm f} = 0.32$ K, marking the lowest temperature of the RS–like state (see Table~\ref{tab:exponents}). This behavior is further corroborated microscopically by the muon spin relaxation rate, $\lambda_{\rm ZF} \sim T^{-0.57}$. Notably, the exponents derived from both thermodynamic and microscopic probes  lie within a similar range, offering strong evidence for the emergence of a RS–like state in the intermediate temperature window between $T_{\rm f}$ and $T^{*}$ in zero field. It is worth noting that within the gigahertz frequency regime, the spin dynamics are dominated by unconventional spin correlations, whereas the inhomogeneous RS-type fluctuations observed at megahertz frequencies remain unresolved (see Table \ref{tab:exponents}).\\\\
We recall that singlets initially form on the shortest and strongest bonds, while spins connected by longer or weaker bonds remain effectively unpaired at higher temperatures. Upon further lowering the temperature, singlet formation extends over increasing length scales, resulting in a reduced fraction of weakly coupled orphan-like local moments that nevertheless dominate the low-energy magnetic response. Importantly, this implies a thermal redistribution of spectral weight between orphan-spin and singlet-spin contributions.\\
To further examine the scenario of RS state in applied magnetic fields, Fig.~\ref{DC} presents the scaling analysis of $\chi(T)$, $M(H)$, and $C_{\rm mag}(T)$, commonly proposed as hallmarks of a disorder-induced RS–like state \cite{Kimchi2018,PhysRevLett.125.117206,PhysRevLett.122.167202}. In Fig.~\ref{DC}(a), the scaled form $\mu_{0}H^{0.56}\chi(T)$ collapses onto a universal curve when plotted against $T/\mu_{0}H$, consistent with RS scaling behaviour. Although the exponent $\alpha_{\chi}$= 0.56 is close to that expected near zero field condition ($\alpha$ =  0.50 to 0.59), the scaling of $T^{-0.45}M$ versus $\mu_{0}H/T$ (Fig.~\ref{DC}(b)) produces a single curve over 2–80 K with complementary exponent ($\alpha_{M} \sim 1 - \alpha_{\chi}$). Furthermore, the low-temperature specific heat measured in various magnetic fields collapses as $\mu_{0}H^{0.25}C_{\rm mag}(T)/T$ vs $T/\mu_{0}H$, yielding an exponent $\alpha_{C}$ = 0.25 even lower than that extracted from zero-field data (Fig.~\ref{DC}(c)). 
These observations suggest that, although scaling collapse is retained under magnetic fields, the exponents differ substantially from their zero-field counterparts. A similar trend is found in the collapse behavior of thermodynamic quantities in Y$_{2}$CuTiO$_{6}$ \cite{PhysRevLett.125.117206}.  Unlike in 1D RS systems, disorder-driven singlet formation in higher dimensional frustrated magnets is expected to exhibit nonuniversal effective exponents. Moreover, both the existence and the universality of a well-defined 3D RS state remain open questions. The 1 T data are intentionally excluded from the scaling plot in Fig.~\ref{DC}(c), as satisfactory data collapse is observed only for higher magnetic fields (3–9 T), similar to what has been reported for the 3D RS candidate Li$_{4}$CuTeO$_{6}$~\cite{Khatua2022}.\\
Although in the intermediate temperature range the system exhibits RS–like state, at lower temperatures around $T_{\rm f} = 0.32$ K it undergoes spin freezing, as revealed by the \textit{ac} magnetic susceptibility measurements. Notably, this freezing occurs only at very low temperatures, suggesting an origin in the Cu$_4$ spin chains, where the Cu$_4$–Cu$4$ coupling ($J_4 = 1.23$ K) drives a quasi-frozen state within the dominant RS background set by the 3D hyper-star lattice spin topology.  Under magnetic fields, however, this quasi-frozen contribution is gradually suppressed, and by 9 T the magnetic specific heat recovers a quadratic dependence, consistent with low-energy gapless excitations~\cite{PhysRevLett.127.157204}.

\section{Conclusion}
In summary, thermodynamic and microscopic measurements identify Li$_2$Cu$_2$(MoO$_4$)$_3$ as a rare 3D hyper-star lattice, where Li/Cu site disorder drives a crossover from  antiferromagnetic short-range correlations at high temperatures to  a RS–like state at low temperatures. The broad  maximum  in $C_{\rm mag}(T)$ and power-law behaviors of the muon relaxation rate and ESR linewidth ($T^{0.26}$) above $T^{*}$ $\approx$ 15.8 indicate dominant short-range correlations characteristic of low-dimensional antiferromagnets. Below $T^{*}$, orphan spins with weak random interactions produce distinct power-law dependences in various physical observables ($T^{-0.50}$) with comparable exponents, evidencing an RS-like state. The universal scaling of $\chi(T)$, $M(H)$, and $C_{\rm mag}(T)$ under field supports this scenario, while varying exponents indicate an intriguing nature of higher-dimensional RS state. Upon further cooling, a quasi-frozen state emerges near $T_{\rm f} = 0.32$ K, likely from weakly coupled Cu$4$ spin chains. Altogether, Li$_2$Cu$_2$(MoO$_4$)$_3$ reveals that site disorder in a 3D frustrated lattice can intertwine bulk and orphan spins, giving rise to an extended RS-like quantum-disordered state beyond two dimensions.

    \section*{Acknowledgments}
    The work at SKKU was supported by the National Research Foundation
  (NRF) of Korea (Grant no. RS-2023-00209121, 2020R1A5A1016518). S.M.K. and C.-L.H. are supported by the
  National Science and Technology Council in Taiwan with a
  Grant No. NSTC 114-2112-M-006-012. R.S. acknowledges the financial support provided
  739 by the Ministry of Science and Technology in Taiwan under
  740 Projects No. NSTC-114-2124-M-001-009 and No. NSTC-
  741 113-2112M001-045-MY3, as well as support from Academia
  742 Sinica for the budget of AS-iMATE11512. We also acknowl-
  743 edge financial support from the Center of Atomic Initiative for
  744 New Materials (AIMat), National Taiwan University, under
  745 Project No. 113L900801. 
   \section{DATA AVAILABILITY}
 The data supporting the findings of this study are available from the corresponding author upon reasonable request. However, the $\mu$SR data presented in this article are openly accessible at Ref.~\cite{psidatabase}.
  \begin{figure}
  	\centering
  	\includegraphics[width=0.45\textwidth]{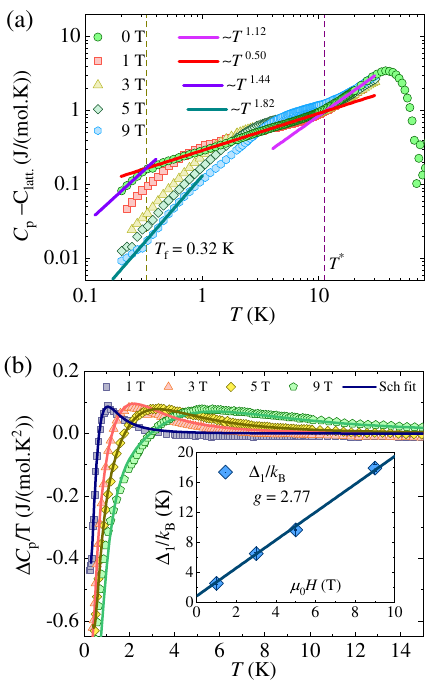}
  	\caption{(a) Temperature dependence of the residual specific heat obtained after subtracting the lattice contribution. In zero field, three distinct power-law regimes are observed, separated by two vertical lines at $T^{*} = 11.11$ K and $T_{\rm f} = 0.32$ K.  (b) Fitting of the Schottky contribution at several magnetic fields. Inset shows the estimated energy gap as a function of magnetic field. }{\label{APPn}}.
  \end{figure} 
   \appendix   
   \section{Specific heat}
 Figure~\ref{APPn}(a) shows the residual specific heat after subtraction of the lattice contribution. The zero-field data are free from any field-induced Schottky contribution, which typically arises under applied magnetic fields due to Zeeman splitting of  $s = 1/2$ orphan spins or spin singlets. In zero field, the data exhibit three distinct power-law regimes in temperature, reflecting successive changes in spin correlations upon cooling. With increasing magnetic field, the narrow low-temperature slope below 0.32 K gradually decreases, indicating a field-induced suppression of the quasi-frozen state and the emergence of an almost quadratic temperature dependence at 9 T, alluding to the field-induced emergence of exotic gapless excitations. \\
 In order to subtract the Schottky contribution for $H \neq 0$, the zero-field data $C_{p}(0~\mathrm{T})$ were subtracted from the corresponding data at finite fields $C_{p}(H \neq 0)$. The resulting $\Delta C_{p}/T$ data are presented in Fig.~\ref{APPn}(b), which were fitted using the function $[C_{\mathrm{Sch}}(\Delta_{1}) - C_{\mathrm{Sch}}(\Delta_{2})]\eta_{\rm orp}/T$, where $\eta_{\rm orp}$ represents the fraction of orphan spins, and $\Delta_{1}$ and $\Delta_{2}$ denote the energy gaps at finite and zero fields, respectively. The Schottky specific heat is given by $C_{\mathrm{Sch}}(\Delta) = R \left( \frac{\Delta}{k_{\mathrm{B}}T} \right)^{2} \frac{\exp\!\left(\frac{\Delta}{k_{\mathrm{B}}T}\right)}{\left[1 + \exp\!\left(\frac{\Delta}{k_{\mathrm{B}}T}\right)\right]^{2}}$,where $k_{\mathrm{B}}$ is the Boltzmann constant.
A similar procedure to subtract the Schottky contribution has been previously applied to materials such as  Y$_2$CuTiO$_{6}$ \cite{PhysRevLett.125.117206} and Ba$_{3}$CuSb$_{2}$O$_{9}$ \cite{PhysRevLett.106.147204}.
As shown in the inset of Fig.~\ref{APPn}(b), $\Delta_{1}/k_{\rm B}$ varies linearly with the applied magnetic field. The linear fit (solid line) yields a $g$-factor of 2.77, slightly higher than the value typically observed for Cu$^{2+}$ ($s = 1/2$) ions~\cite{PhysRevLett.125.117206,PhysRevLett.106.147204}. This discrepancy suggests that the orphan spins are dressed by their coupling to the surrounding RS environment. Thus, the extracted gap reflects the weighted average of the local environment. At higher fields, the dominance of the high-energy tail of the gap distribution leads to an apparent $g$-factor that exceeds the ESR value. \\
 On the other hand, $\Delta_{2}/k_{\rm B}$ increases from 0.89~K to 1.7~K as the magnetic field is raised from 1 to 9~T. This gap is present already in zero field and can be attributed to  spin dimers with varying gaps, which are an essential ingredient of the RS state~\cite{PhysRevLett.125.117206,PhysRevLett.106.147204}.
 After the subtraction of the Schottky contribution from the residual specific heat, the final magnetic specific heat ($C_{\mathrm{mag}}(T)$) is presented in Fig.~\ref{HCFig}(c).\\
Note that the fraction of orphan spins estimated from the Schottky analysis is 3.7\%, 7.01\%, 8.38\%, and 13\% for applied magnetic fields of 1, 3, 5, and 9~T, respectively. The increasing concentration of orphan spins should not be literally interpreted as these spins are bound to the RS background. Rather, it reflects that an increasingly larger number of spins contribute to the observed Schottky behavior with increasing magnetic field.

\bibliography{LCMO}
\end{document}